\documentclass{elsart}

\usepackage{epsfig}

\usepackage{amssymb}

\def\ls{{_<\atop^{\sim}}}
\def\gs{{_>\atop^{\sim}}}

\begin{document}

\begin{frontmatter}



\title{Measured Cosmological Mass Density in the WHIM: the Solution to 
the 'Missing Baryons' Problem}


\author[cfa]{Fabrizio Nicastro\corauthref{cor1}}
\ead{fnicastro@cfa.harvard.edu}
\author[cfa]{Martin Elvis}
\author[oar]{Fabrizio Fiore}
\author[osu]{Smita Mathur}
\address[cfa]{Harvard-Smithsonian Center for Astrophysics, 60 Garden Street, 
MS-83, Cambridge, MA, 02138, U.S.A.}
\address[osu]{Astronomy Department, The Ohio State University, 43210, 
Columbus, OH, U.S.A.}
\address[oar]{Osservatorio Astronomico di Roma, Monteporzio Catone, Italy}
 
\corauth[cor1]{F. Nicastro}

\begin{abstract}
We review the current high-significance X-ray detections of Warm-Hot 
Intergalactic Medium (WHIM) filaments at $z > 0$ along the lines of sight 
to the two blazars Mrk~421 ($z=0.03$) and 1ES~1028+511 ($z=0.361$). 
For these WHIM 
filaments, we derive ionization corrections and, when possible, metallicity 
estimates. This allows us to obtain refined estimates of the 
number density of O\,VII WHIM systems down to the O\,VII column density 
sensitivity of our observations, and most importantly, a measurement of the 
cosmological mass density $\Omega_{\mathrm{b}}^{\mathrm{WHIM}}$ in the WHIM, at redshift 
$z<0.361$. These estimates agree well with model predictions and with the 
total estimated amount of missing baryons in the local Universe, although 
errors are large, due to the still limited number of systems. We conclude 
discussing future observational strategies and mission designs for WHIM 
studies. 
\end{abstract}

\end{frontmatter}

\section{'Missing Baryons' and the Warm-Hot Intergalactic Medium}
\label{WHIM}
The total number of baryons in the Universe, as inferred by both Big-Bang 
Nucleosynthesis (BBN) compared with observations of light element 
ratios (Kirkman et al., 2003) and Cosmic Microwave Background (CMB) 
anisotropies Bennett et al., 2003; Spergel et al., 2003) 
should amount to $\Omega_{\mathrm{b}} = (4.6 \pm 0.2)$ \% of the total cosmological 
mass/energy density $\Omega_{\mathrm{b}} + \Omega_{\mathrm{DM}} + \Omega_{\mathrm{DE}}$. 
While this number is in good agreement with the number of baryons 'counted' 
in the H\,I Ly$\alpha$ Forest at $z > 2$, $\Omega_{\mathrm{b}} > 3.5$ \% (Rauch, 1998; 
Weinberg et al., 1997), 
a large discrepancy is found in the Local Universe, at $z < 2$. The total 
number of baryons seen at low redshift in virialized matter (i.e. 
stars, neutral H and He, molecular H, and X-ray emitting plasma in clusters 
of galaxies: Fukugita, 2003) and not yet virialized matter (residual H\,I 
Ly$\alpha$ Forest - Penton, Stocke \& Shull, 2004 -, and both photoionized 
and collisionally ionized O\,VI in the Intergalactic Medium - Tripp, Savage \& 
Jenkins, 2000; Savage et al., 2002) amount to only $\Omega_{\mathrm{b}}(z < 2) = (2.5 
\pm 0.3)$ \%. 
About 46 \% of the total predicted baryons are then missing at 
$z < 2$: $\Omega_{\mathrm{b}}(\mathrm{missing}) = (2.1^{+0.5}_{-0.4})$ \% (Nicastro et al., 
2004a - N04a -, and references therein). 

Hydrodynamical simulations for the formation 
of large-scale structures in the Universe predict that these baryons 
hide in a filamentary web of tenuous ($n_{\mathrm{b}} \simeq 10^{-6}-5 \times 10^{-5}$ 
cm$^{-3}$, i.e. $\delta = n_{\mathrm{b}}/<n_{\mathrm{b}}> \simeq 5-100$
\footnote{$<n_{\mathrm{b}}> = 2 \times 10^{-7} (1+z)^3 (\Omega_{\mathrm{b}}h^2/0.02)$ is the 
average density in the Universe.}
) matter at warm-hot temperatures ($T \simeq 10^5-10^7$ K) that permeates 
intergalactic space and connects already virialized structures (e.g. 
Hellsten, Gnedin \& Miralda-Escud\'e, 1998; Cen \& Ostriker, 1999; Dav\'e 
et al., 2002; Fang, Bryan \& Canizares, 2002). 
This primordial matter would provide the fuel for the assembly of galaxies 
and galaxy groups and clusters, and would be shock-heated 
during the continuous infall process of structure formation. 
The internal shocks 
ionizes the gas to such a degree that it becomes 'invisible' in 
infrared, optical or UV light, but should shine and absorb Far-UV and 
X-ray photons. However, given the very low-density, extremely high 
sensitivity, large-field of view, FUV or X-ray detectors are needed 
to image the emission from these filaments (e.g. Yoshikawa et al., 2003), 
and these are not 
yet available. The only avenue pursuable today, is to detect the WHIM in 
absorption against bright background sources. 

Here we present two LETG spectra of the blazars Mrk~421 and 1ES~1028+511, and the 
first detections 
of three O\,VII WHIM filaments. A fourth filament is only 
detected in C\,V$_{1s\rightarrow2p}$ in the 1ES~1028+511 spectrum. 
Sections 4 and 5 are devoted to the estimates of the number density and the 
cosmological mass density in the O\,VII WHIM, while in section 6 we discuss 
future prospects for WHIM studies. 
Throughout the paper we use LETG spectra grouped at a resolution of 
$\Delta \lambda = 12.5$ m\AA\ for fitting purposes, i.e. 4 times better 
than the intrinsic LETG resolution of 50 m\AA. The plots, instead, show 
spectra binned at 25 m\AA\ for Mrk~421 and 12.5 m\AA\ for 1ES~1028+511. 
Equivalent Width sensitivity limits are converted into ion column density 
limits by using Doppler parameters corresponding to thermal broadening of the 
lines in plasma at logT(K) = 6.1, the central temperature of the WHIM 
distribution (e.g. Dav\'e et al., 2002). 
We adopt $H_0 = 72$ km s$^{-1}$ Mpc$^{-1}$ (Freedman et al., 2001, 
Bennett et al., 2003). 
Errors are quoted at 1$\sigma$ for 1 interesting parameter, unless 
otherwise stated. 

\section{LETG Spectra of Mrk~421 and 1ES~1028+511}
{\bf Mrk~421} is a blazar at $z=0.03$. 
We observed Mrk~421 with the {\em Chandra} LETG during two very 
luminous outbursts, on 2002, October, 26-27 and 2003, July 1-2, under our 
TOO programs to observe blazars in outburst phases. 
These two observations lasted $\sim 100$ ks each and caught the source at 
historical maxima, 60 and 40 mCrab in the 0.5-2 keV band, allowing us to 
collect a total of $\sim 5300$ Counts Per Resolution Element (CPREs) at 21 
\AA\ (Nicastro et al., 2004b: N04b). This S/N was enough to 
detect O\,VII columns of N$_{\mathrm{O\,VII}} \gs 8 \times 10^{14}$ cm$^{-2}$ at $\ge 
3\sigma$. 
We detected 24 absorption lines in the LETG spectrum of Mrk~421, between 
10 and 50 \AA\ (see No4b for a full, detailed spectral analysis of both 
continuum and lines). 
While the majority of these lines are imprinted by both neutral and highly 
ionized systems at velocity consistent with zero (ISM gas, and either Local 
Group WHIM - Nicastro et al., 2002; Nicastro et al., 2003; Williams et al., 2004 - or 
an extended Galactic Corona - Sembach et al., 2003; Savage et al., 2004), 9 lines are 
instead identified with two 
intervening WHIM systems at $z = (0.011 \pm 0.001)$ and $z = (0.027 \pm 
0.001)$. Figure 1 shows the 21-22.5 \AA\ portion of the LETG spectrum of 
Mrk~421, in which the two O\,VII$_{1s\rightarrow2p}$ lines from the two WHIM systems, 
together with the $z=0$ lines of O\,VI and O\,VII (and possibly a fourth 
O\,VII$_{1s\rightarrow2p}$ transition at $z=0.033$, where also a H\,I Ly$\alpha$ is seen 
- Dull et al., 1996), are present. 
%
\begin{figure}
\epsfysize=2.5in 
\epsfxsize=3.5in 
\hspace{0.8in}
\epsfbox{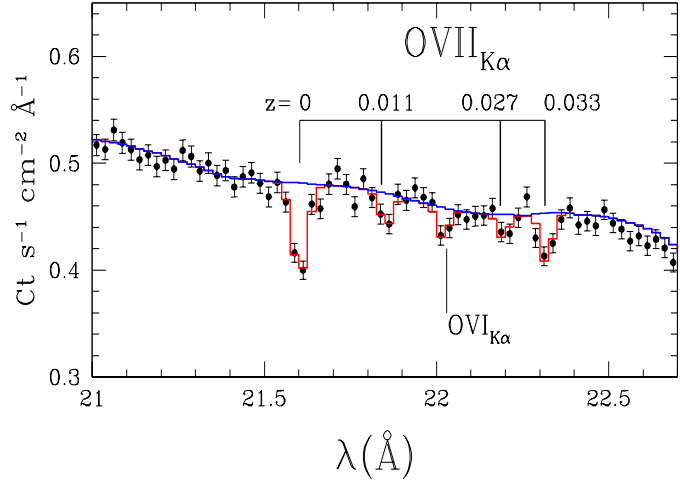}
\vspace{0in}\caption[h]{\footnotesize 21-22.5 \AA\ portion of the 
{\em Chandra}-LETG spectrum of Mrk~421. The dashed line shows the best 
fitting continuum, while the solid line is our best-fitting continuum 
plus line model.}
\end{figure}
%
\begin{figure}
\epsfysize=2.5in 
\epsfxsize=3in 
\hspace{0.8in}
\epsfbox{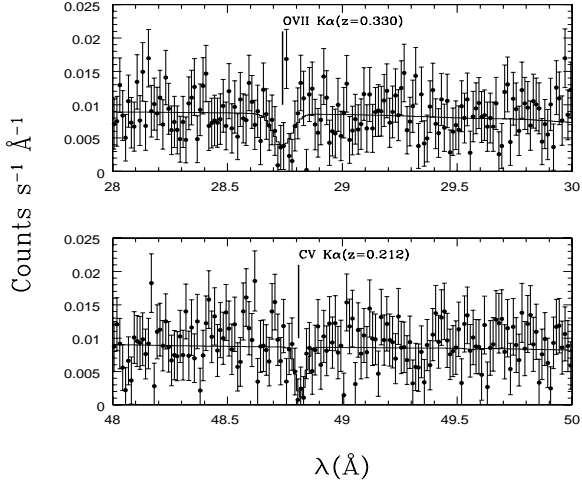}
\vspace{0in}\caption[h]{\footnotesize Two portions of the HRC-LETG spectrum 
of 1ES~1028+511: (a) 28-30 \AA\ and (b) 48-50 \AA. The solid lines show our 
best fit models (the continuum is fitted over the broad range 10-60 \AA). 
Two lines from O\,VII (top panel) and C\,V (bottom panel), from two intervening 
WHIM systems, are detected in these two portions of the 1ES~1028+511 spectrum.}
\end{figure}

{\bf 1ES~1028+511} is a blazar at $z=0.361$. 
It was observed under our {\em Chandra} TOO program, with the 
{\em Chandra} LETG, following a short outburst phase (3-days), on 2004, 
March 11. Although 1ES~1028+511 was back close to its 'quiescent' level 
when {\em Chandra} pointed at it, 1ES~1028+511 turned out to be a bright 
soft X-ray source for such a high redshift ($F_{0.5-2 \mathrm{keV}} = 1.1$ 
mCrab). 
The 150 ks HRC-LETG spectrum of 1ES~1028+511 contains 80-100 CPREs in the 
$21 < \lambda($\AA$) \le 50$ interval, enough to detect at 3$\sigma$ 
N$_{\mathrm{O\,VII}} \ge 1.4 \times 10^{16} (1+z)^{-2}$ cm$^{-2}$ at 
$z \le 0.361$, and 4.1 times lower C\,V columns, N$_{\mathrm{C\,V}} \ge 3.4 \times 10^{15} 
(1+z)^{-2}$ cm$^{-2}$ at $z \le 0.269$ (at $\lambda > 51$ \AA\, i.e. $z > 
0.269$ for C\,V$_{1s\rightarrow2p}$, the spectrum is heavily 
affected by the presence of the HRC plate gap). Although the analysis of 
these data is still in progress, we show here some preliminary results. 
We detect two strong lines (3.5$\sigma$ and 4.4$\sigma$) at $\lambda = 
28.740 \pm 0.038$ \AA\ and $\lambda = 48.816 \pm 0.040$ \AA\, that we 
identify as O\,VII$_{1s\rightarrow2p}$ from a WHIM system at $z=0.330 \pm 0.002$, and 
C\,V$_{1s\rightarrow2p}$ from a WHIM 
system at $z=0.212 \pm 0.001$ (Figure 2 (a) and (b) respectively; Nicastro 
et al., 2005, in preparation: N05). 
Errors on $\lambda$ and $z$ include a 20 m\AA\ 
(1$\sigma$) calibration uncertainty in the HRC-LETG dispersion relationship 
(see e.g. N04b and references therein). We also detect the N\,VII Ly$\alpha$ at 
the redshift of the O\,VII system ($z=0.330$), but only at a significance of 
2.6$\sigma$.  

\section{WHIM Solutions for the O\,VII Systems} 
For the two O\,VII-N\,VII absorbers along the line of sight to Mrk~421 
(and also based on their measured H\,I and O\,VI upper limits, from HST-STIS 
and FUSE respectively - N04b) 
we find self-consistent ionization balance and metallicity solution, 
that allow us to derive equivalent H column density and temperature 
range (N04a, N04b). These solutions are found by comparing the data with 
hybrid models of tenuous (i.e. $n_{\mathrm{e}} \ls 10^{-4}$ cm$^{-3}$) collisionally ionized 
gas, undergoing the additional contribution of photoionization by the diffuse 
extragalactic UV-X-ray background (plus the proximity effect of the beamed radiation 
from the blazar, in the case of the line of sight to Mrk~421: see N04b for details). 
From these solutions, and assuming homogeneity in the systems (which may not be 
appropriate, given the evidence for the co-existence of multi-phase 
structures for at least one of the two filaments: N04b), we derive 
the depth of these filaments along our line of sight, in terms of the 
baryon volume density in units of $n_{\mathrm{b}} = 10^{-5}$ cm$^{-3}$. 
The first two rows of table 1 summarize the results. 
%
\begin{table}
\begin{center}
\caption{\bf \small WHIM Solutions for the Three O\,VII Systems} 
\vspace{0.4truecm}
\begin{tabular}{|ccccc|}
\hline
Redshift & $T$ (10$^6$ K) & N$_{\mathrm{b}}$ $^a$ & D$^b$ & [O/H] \\
\hline
$(0.011 \pm 0.001)$ & (0.6-2.5) & $(2.0 \pm 0.3)$$^c$ & 
$(0.7 \pm 0.1)$$^c$ & $> -1.47$ \\
$(0.027 \pm 0.001)$ & (1.1-1.7) & $(2.8 \pm 0.2)$$^c$ & 
$(0.9 \pm 0.1)$$^c$ & $> -1.32$ \\
\hline
$(0.330 \pm 0.002)$ & (1.6-2.5) & $(24 \pm 8)$$^d$ & 
$(7.8 \pm 2.5)$$^d$ & N/A \\
\hline 
\end{tabular}
\end{center}
\vspace{0.2truecm}
$^a$ In units of $10^{19} 10^{-[O/H]_{-1}}$ cm$^{-2}$. 
$^b$ In units of $10^{-[O/H]_{-1}} (n_{\mathrm{b-5}})^{-1}$ Mpc: 
$(n_{\mathrm{b-5}})$ is the baryon density in units of $10^{-5}$ cm$^{-3}$. 
$^c$ At logT = 6.1. 
$^d$ At logT = 6.3. 
\end{table}
%

For the two O\,VII/C\,V systems along the line of sight to 1ES~1028+511, 
our analysis is still in progress. Table 1 (3rd row) contains only rough 
estimates of the equivalent H column density, temperature and depth of the 
O\,VII filament, derived based on the N$_{\mathrm{O\,VII}}$, N$_{\mathrm{N\,VII}}$ measurements and 
the N$_{\mathrm{O\,VIII}}$ 3$\sigma$ upper limit. Metallicity estimates for this object 
are currently not available. 

\section{Number Density of O\,VII WHIM Filaments}
Figure 4 shows the predicted cumulative number of O\,VII WHIM system per unit 
redshift as a function of the minimum O\,VII column density 
(solid curve; Fang, Bryan \& Canizares, 2002). The lowest O\,VII column density 
we detect along 
the line of sight to Mrk~421, is N$_{\mathrm{O\,VII}} = 7 \times 10^{14}$ cm$^{-2}$, 
which is about the 3$\sigma$ sensitivity of the spectrum of Mrk~421 in the 
entire $z \le 0.03$ redshift range.  
The redshift of 1ES~1028+511, instead, is much higher, so we cannot 
neglect the dependence of the N$_{\mathrm{O\,VII}}$ detection sensitivity threshold 
on the wavelength (see \S 6). 
The LETG spectrum of 1ES~1028+511 has a $> 3\sigma$-sensitivity to 
N$_{\mathrm{O\,VII}} = 8.7 \times 10^{15}$ cm$^{-2}$ (the column of the 
detected O\,VII line at $\lambda = 28.74$) for $0.269 \le z \le 0.361$, a 
redshift interval of $\Delta z = 0.092$. 
At $z < 0.269$ 
the effective redshift range within which we can detect N$_{\mathrm{O\,VII}} \ge 
8.7 \times 10^{15}$ cm$^{-2}$ at a significance $> 3\sigma$ is 
$ \Delta z_{\mathrm{eff}} = 0.075$ (see N05 for details). 
So, the total effective redshift range is $\Delta z_{\mathrm{tot}} = \Delta 
z_{\mathrm{eff}} + \Delta z = 0.167$. 

The cumulative numbers of $> 3\sigma$ detected O\,VII WHIM filaments per unit 
redshifts, down to N$_{\mathrm{O\,VII}}^{\mathrm{Thres}} = 7 \times 10^{14}$ cm$^{-2}$ and 
N$_{\mathrm{O\,VII}}^{\mathrm{Thres}} = 8.7 \times 10^{15}$ cm$^{-2}$, are therefore: 
$d\mathcal{N}/dz(>7\times 10^{14}) = 70^{+90}_{-40}$ and $d\mathcal{N}/dz(>8.7
\times 10^{15}) = 6^{+13}_{-5}$ (Fig. 3; the large errors are due 
to the small number statistics: Gehrels, 1986). 
At 1.4$\sigma$ the two points are inconsistent with a constant value. 
The best fit slope of the power law connecting the two points is $\alpha = 
1.0^{+0.5}_{-0.7}$. 
The solid line in Figure 3 shows that the model predictions of [14] are
consistent (within the large errors) in both normalization and slope with 
the observations. 
%
\begin{figure}
\epsfysize=2.5in 
\epsfxsize=3in 
\hspace{0.8in}
\epsfbox{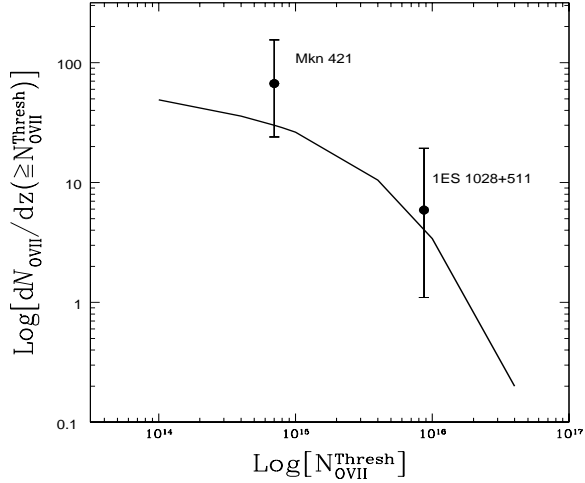}
\vspace{0in}\caption[h]{\footnotesize Predicted cumulative number of O\,VII 
WHIM systems per unit redshift as a function of the minimum O\,VII column 
density (solid curve; Fang, Bryan \& Canizares, 2002). The two points are the 
cumulative numbers of 
$> 3\sigma$ detected O\,VII WHIM filaments per unit redshifts, down to 
N$_{\mathrm{O\,VII}}^{\mathrm{Thres}} = 7 \times 10^{14}$ cm$^{-2}$ (line of sight to Mrk~421) 
and N$_{\mathrm{O\,VII}}^{\mathrm{Thres}} = 8.7 \times 10^{15}$ cm$^{-2}$ (line of sight to 
1ES~1028+511).}
\end{figure}

\section{Cosmological Mass Density of O\,VII WHIM Filaments}
Based on the equivalent H column density measurements from 
three O\,VII WHIM filaments (Table 1), we can now 
estimate the baryonic cosmological mass densities $\Omega_{\mathrm{b}}^{\mathrm{WHIM}}$ 
down to our two values of N$_{\mathrm{O\,VII}}^{\mathrm{Thres}}$. 
We obtain $\Omega_{\mathrm{b}}^{\mathrm{WHIM}}(>7\times 
10^{14}) = (2.7^{+3.8}_{-0.9}) \times 10^{-[O/H]_{-1}}$ \% (N04a) and 
$\Omega_{\mathrm{b}}^{\mathrm{WHIM}}(>8.7\times 10^{15}) = (2.0^{+4.7}_{-1.8}) \times 
10^{-[O/H]_{-1}}$ \% (N05), for the lines of sight to Mrk~421 and 1ES~1028+511 
respectively. Errors on these estimates include the (dominant) 
asymmetric Poisson errors associated with a small number of events (Gehrels, 1986). 
One can try to reduce these errors combining the two $\Omega_{\mathrm{b}}^{\mathrm{WHIM}}$ 
measurements. 
However, to do this, we have to extrapolate the 1ES~1028+511 measurement, 
down to the sensitivity of the Mrk~421 spectrum, i.e. down to N$_{\mathrm{O\,VII}} = 7 
\times 10^{14}$ cm$^{-2}$. First of all we have to derive 
an equivalent redshift range $\Delta z_{\mathrm{eq}}$, within which we would expect 
to detect a single O\,VII WHIM filament with N$_{\mathrm{O\,VII}} = 7 \times 10^{14}$ 
cm$^{-2}$ along the line of sight to 1ES~1028+511. 
We obtain $\Delta z_{\mathrm{eq}} = 
0.023$ (see N05 for details). 
From $\Delta z_{\mathrm{eq}}$ we can derive an equivalent $d_{\mathrm{eq}}^{\mathrm{1 ES}}$ 
distance to 1ES~1028+511 that we will use in the formula for $\Omega_{\mathrm{b}}$. 
Next we need to evaluate the probability $P$ that this single O\,VII filament 
has N$_{\mathrm{O\,VII}} \ge 8.7 \times 10^{15}$ cm$^{-2}$, and use it in the 
formula for $\Omega_{\mathrm{b}}$ to weight the N$_{\mathrm{H}}^{\mathrm{1 ES}}$ measurement (Table 1). 
We find $P 
\simeq 0.14$ (see N05 for details). 
This gives a combined $\Omega_{\mathrm{b}}$ estimate 
$\Omega_{\mathrm{b}}^{\mathrm{WHIM}}(\ge 7 \times 10^{14}) = (2.4^{+1.9}_{-1.1}) \times 
10^{-[O/H]_{-1}}$ \%, consistent with both model predictions and the 
actual number of missing baryons. 

\section{WHIM Detectability}
The WHIM imprints high-ionization metal absorption 
lines, both in the FUV (mainly O\,VI) and in the X-rays (transitions from 
He-like and H-like C, N, O and Ne, and inner shell transitions from lower 
ionization ions). 
While high signal-to-noise FUSE FUV spectra of tens 
of low-redshift AGNs are readily available, the observed number density 
of relatively cool ($T \simeq 10^5-10^{5.5}$ K) shock-heated O\,VI 
filaments with N$_{\mathrm{O\,VI}} \ge 4 \times 10^{13}$ cm$^{-2}$ (the average 
threshold sensitivity of high signal to noise FUSE spectra of AGNs) 
is low: $d\mathcal{N}/dz = 14^{+9}_{-6}$ (Savage et al., 2002). 
The predicted number density of the hotter ($T \simeq 10^{5.5}-10^{6.5}$) 
O\,VII-dominated filaments, down to the same columns (N$_{\mathrm{O\,VII}} \ge 4 \times 
10^{13}$ cm$^{-2}$), is a factor of 5 larger, and decreases to that observed 
for O\,VI at column densities $\sim 2$ orders of magnitude larger (N$_{\mathrm{O\,VII}} 
\ge 4 \times 10^{15}$ cm$^{-2}$: Fig. 3). 
The total equivalent H column density (and so the baryon mass) implied by 
'pure' O\,VII filaments (i.e. systems in which the O\,VI columns are below the 
current sensitivity threshold) is then more than one order of magnitude larger 
than that of 'pure' O\,VI filaments: the bulk of the WHIM can only be detected 
in the X-ray band. 
%
\begin{figure}
\epsfysize=2.5in 
\epsfxsize=3in 
\hspace{0.8in}
\epsfbox{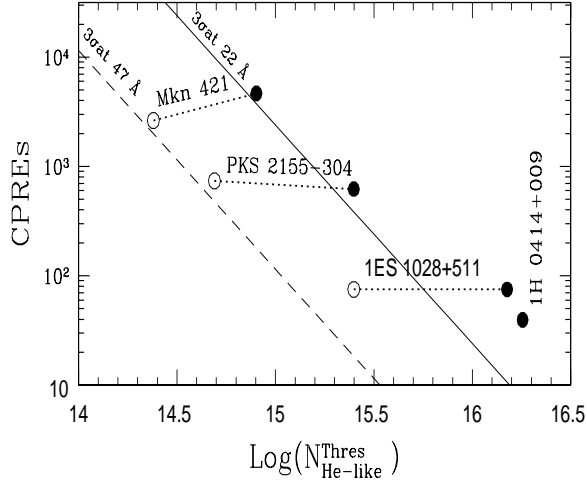}
\vspace{0in}\caption[h]{\footnotesize CPREs as a function of 3$\sigma$ 
N$_{\mathrm{He-like}}^{\mathrm{Thres}}$ (we use the oscillator strength of the O\,VII$_{1s\rightarrow2p}$, 
$f=0.696$). The solid and dashed lines are the approximate analytical 
relation N$_{\mathrm{He-like}}^{\mathrm{Thres}} \simeq 1.1 \times 10^{17} N_{\sigma} 
(\Delta\lambda($m\AA$)) ($CPREs$)^{-1/2} \lambda($\AA$)^{-2}$, at 22 \AA\ and 47 \AA, 
respectively. In this formula columns are in cm$^{-2}$, $N_{\sigma}$ is the 
number of $\sigma$ (3 in the plotted curves), $\Delta\lambda$ is the grating 
resolution in m\AA\ (50 m\AA\ for both curves). 
Data points are the measured 3$\sigma$ N$_{/mathrm{He-like}}^{\mathrm{Thres}}$ upper limits at 
22 \AA\ (filled circles) and at 47 \AA\ (empty circles) from the 
{\em Chandra} spectra of Mrk~421, PKS~2155-304 and 1ES~1028+511 and 
the XMM-{\em Newton} spectrum of 1H~0414+009. 
These empirical data nicely confirm the analytical 
curves for CPRE$\gs 100$. At lower CPREs data deviate from 
the ideal analytical curves, due to deviations of the actual 
signal-to-noise ratio in the data from the ideal Poissonian $\sqrt{\mathrm{CPREs}}$. 
We note that about 80 CPREs are needed for a 3$\sigma$ detection of 
N$_{\mathrm{He-like}} = 1.5 \times 10^{16}$ cm$^{-2}$ at 22 \AA\ (1ES~1028+511,  
filled circle), but about 6 times lower columns are $3\sigma$-detectable 
with the same number of CPRE, at 47 \AA\ (1ES~1028+511, empty circle).}
\end{figure}
Detecting He-like absorption 
lines with N$_{\mathrm{ion}} \le 2 \times 10^{15}$ at $\lambda = 
22$ \AA\ (about the restframe wavelength of O\,VII$_{1s\rightarrow2p}$) at the 
spectral resolution of the current X-ray spectrometers 
($\Delta \lambda = 50$ m\AA\ for both the {\em Chandra} Low Energy 
Transmission Grating - LETG - and the XMM-{\em Newton} Reflection Grating 
Spectrometer - RGS), requires X-ray spectra with $\ge 600$ CPREs in the continuum 
(Fig. 4, solid line). 
This requirement relaxes significantly at longer wavelength, since 
N$_{\mathrm{ion}}^{Thres}$ (for given grating resolution 
and CPREs) decreases with $\lambda^{-2}$. 
So, for example, at $\lambda = 47$ \AA\ ($z = 1.18$ for the O\,VII$_{1s\rightarrow2p}$, 
but only $z=0.16$ for the C\,V$_{1s\rightarrow2p}$) only $\sim 120$ CPREs are required to 
reach the same 3$\sigma$ sensitivity of N$_{\mathrm{ion}} \le 2 \times 10^{15}$ (Fig. 
4, dashed line). 
Unfortunately, however, the highest quality LETG or RGS spectra of the 
brightest (and very nearby) AGNs contains at most 50 CPREs. Two important 
exceptions are the cases presented here: the LETG spectra of the two 
blazars Mrk~421 ($z=0.03$) and 1ES~1028+511 ($z=0.361$), observed by 
{\em Chandra} under our TOO programs, while undergoing outbursts. 

\section{Future Prospects}
This work only begins the study of the Warm-Hot IGM outside the Local 
group (Nicastro et al., 2002, 2003; Williams et al., 2004), now that the WHIM has 
been detected and theoretical predictions have been verified. 

The large number of ion species showing up in the X-ray spectra 
imply a rich field of investigation. 
Higher significance detections, along several 
other lines of sight, and at higher redshift, of He-like and H-like 
transitions from C, N, O and Ne, will allow us to measure the relative 
metallicity ratios at different redshifts. 
This will fundamentally contribute to the assessment of the still poor 
measurements of metal production with cosmic age, and will help 
refining mechanisms for galaxy/AGN-IGM feedback (e.g. galaxy 
superwinds, quasar winds). 
Moreover, the potential for the simultaneous detection, in the same band, of 
inner-shell transition from lower ionization species of the same ions, will 
allow us to distinguish between different ionization scenarios, and to assess 
the importance and frequency of multiphase IGM. Spectral resolutions of 
$R \ge 3000$ in the soft X-ray band will allow us to separate different 
kinematics components. This has already been proven very useful in the FUV 
band, for example to distinguish between a Local Group WHIM origin and a halo 
origin for the O\,VI high-velocity absorbers at $z \simeq 0$ (Nicastro et al., 2003). 
Reconstructing the distribution 
functions of several ions in the WHIM (i.e. $d\mathcal{N}/dzdN$), will also 
probe very useful in distinguishing different cosmologies, and possibly 
tracing the distribution of dark-matter potential wells in the local Universe. 

Observing sources with {\em Chandra} while in bright outburst phases has 
proven fruitful, allowing the detection of the WHIM for the first time. 
However, Mrk~421 is unique: no other extragalactic sources in the 
sky (except Gamma-Ray Burst X-ray afterglows, e.g. Fiore et al., 2000) reach 
the level of Mrk~421 during the observations presented here. While 
relatively short {\em Chandra} and or XMM-{\em Newton} observations 
of blazars in outburst will continue to provide a 
'cheap' way to probe new sightlines for WHIM studies, substantial 
progresses in our understanding of the WHIM phenomenon can only be made by 
exploiting  different observational strategies. 

\noindent 
As shown in Fig. 3, the expected number of O\,VII WHIM filaments 
down to a given N$_{\mathrm{O\,VII}}$ threshold, increases linearly with redshift. About 
2 O\,VII filaments with N$_{\mathrm{O\,VII}} \ge 10^{16}$ are expected up to $z=0.5$ along 
a random line of sight. Such O\,VII columns can easily be detected with 100 ks 
{\em Chandra} and/or {\em Newton}-XMM observations of background sources 
flaring at a level even 10 times lower than that reached by Mrk~421 during 
the observations presented here. 

An important metric for future instrumentation to explore the WHIM,
as shown in figure 4, is that the sensitivity of a grating spectrum (i.e. 
$\Delta\lambda = const$) with constant signal to noise per resolution 
element, to a given column density, increases rapidly with redshift 
($\propto (1+z)^2$). 
Smaller columns are detectable at higher redshift, as beautifully proven by 
the faint C\,V$_{1s\rightarrow2p}$ detection in the moderate signal-to-noise {\em Chandra} 
spectrum of 1ES~1028+511 (about 60 times fainter flux than the {\em 
Chandra} Mrk~421 observations presented here: Fig. 2b). 
These are key elements for planning future observational strategies and 
mission designs for WHIM studies. 
Long integrations, $\gs 0.5-1$ Ms, with {\em Chandra} 
and/or XMM-{\em Newton} of high-$z$ sources in their quiescent states are 
low-risk and possibly the only way to 
dramatically increase the number of detections and so narrow down the large 
statistical errors on the $\Omega_{\mathrm{b}}^{\mathrm{WHIM}}$ estimate. Tripling the currently 
detected number of WHIM systems, by exploiting the above observational 
strategy, is within the reach of current X-ray instruments, and will allow us 
to reduce the uncertainties on $\Omega_{\mathrm{b}}^{\mathrm{WHIM}}$ down to $\sim ^{+60}_{-40}$ 
\%, the level to which the other baryonic components in the local Universe 
are now known (e.g. N04a and reference therein). 

Further advances in this direction will have to wait for the large 
collecting area of the spectrometers on the two planned X-ray missions, 
{\em Constellation}-X
\footnote{http://constellation.gsfc.nasa.gov/science/index.html}
and {\em XEUS}
\footnote{http://www.rssd.esa.int/index.php?project=XEUS}
or dedicated small missions, e.g. {\em Pharos} (Elvis et al., 2002). 
With $\sim$100 times larger effective area and $R>$1000, {\em Constellation}-X 
will allow us to detect WHIM systems as weak as those presented in this work, 
for dozens of different lines of sight, and thicker filaments 
will be detected towards hundreds of different sightlines. 

This will be much more effectively done with gratings than calorimeters, 
particularly if the currently undergoing tests on the off-plane reflection 
gratings configuration for {\em Constellation-X} will confirm the recently 
reached first-order peak absolute efficiency of 43 \% (Heilmann et al., 2004)
\footnote{snl.mit.edu/papers/papers/2003/Heilmann/RKH-SPIE-5168.pdf}
. In calorimeters the resolution is constant in energy, and so the 
resolution element in $\Delta\lambda$ degrades as $\propto \lambda^2$. 
This dependence exactly cancels out the increase in N$_{\mathrm{ion}}$ sensitivity 
given by the formula N$_{\mathrm{ion}}^{\mathrm{Thres}} \gs (\Delta\lambda) \lambda^{-2}$, 
at $\Delta\lambda = const$ (Fig. 4). 
Moreover, the degradation of calorimeter resolution with wavelength 
would greatly hamper the possibility of resolving WHIM lines (expected at 
$\lambda > 20$ \AA) or detecting multiphase structures in the IGM (and so 
prevent tests of the current paradigms for the dominant heating mechanism: 
internal shocks). 
This would in turn hamper our capabilities of tightly constrain the physical 
properties of the baryons in this diffuse web of WHIM. 

A resolving power R$\gs 5000$ is needed to resolve thermally broadened O lines 
in gas at $T=10^6$ K and so to allow for separation of different components, 
and for a clear physical diagnostics (Elvis et al., 2002), which may be achievable 
with {\em Constellation}-X or {\em Pharos}. 

Finally we note that it is crucial that the intrinsic noise of 
the detector containing the dispersed spectrum be extremely low. 
Figure 4 shows that the lowest CPRE point, from the RGS observation of 
1H~0414+009, is actually closer to the ideal analytical sensitivity curve, 
CPREs vs N$_{\mathrm{He-like}}^{\mathrm{Thres}}$, than the one from the HRC-LETG 
observation of 1ES~1028+511. 
The HRC-LETG spectrum of 1ES~1028+511 has almost twice as 
many CPRE as the RGS spectrum of 1H~0414+009, but a 5 times larger 
intrinsic instrumental noise ($5.2 \times 10^{-5}$ cts s$^{-1}$ per 
resolution element), which greatly penalizes the detection of 
faint columns. 
High resolving-power/efficiency X-ray gratings dispersed onto low-noise detectors are 
therefore, by far, the most efficient instruments to systematically study the 
WHIM in the near future. 

\thebibliography{}
\item{} Bennet, C.L. et al., First Year Wilkinson Microwave Anisotropy Probe 
(WMAP) Observations: Preliminary Maps and Basic Results, ApJS, 148, 97-117 
(2003) 
\item{} Cen, R. \& Ostriker, J.P., Where are the Baryons?, ApJ, 514, 1-6 
(1999) 
\item{} Dav\'e, R. et al., Baryons in the Warm-Hot Intergalactic Medium, 
ApJ, 564, 604-623 (2002) 
\item{} Elvis, M., Fiore, F. \& the Pharos Team, A High Resolution 
Intergalactic Explorer for the Soft X-ray/FUV, SPIE (August 2002), 
astro-ph/0303444 (2003) 
\item{} Fang, T., Bryan, G.L. \& Canizares, C.R., Simulating the X-ray 
Forest, ApJ, 564, 604-623 (2002) 
\item{} Fiore, F., et al., Probing the Warm Intergalactic Medium through 
Absorption against Gamma-Ray Burst X-Ray Afterglows, ApJ, 544, L7-L10 
(2000) 
\item{} Fukugita, M., Cosmic Matter Distribution: Cosmic Budget Revisited, 
to be published in the proceedings of IAU symposium 220, Dark Matter in 
Galaxies, astro-ph/0312517 (2003) 
\item{} Gehrels, N., Confidence Limits for Small Numbers of Events in 
Astrophysical Data, ApJ, 303, 336-346 (1986) 
\item{} Hellsten, U., Gnedin, N.Y. \& Miralda-Escud\'e, J., 
The X-Ray Forest: A New Prediction of Hierarchical Structure Formation Models, 
ApJ, 509, 56-61 (1998) 
\item{} Kirkman, D. et al., The Cosmological Baryon Density from the 
Deuterium-to-Hydrogen Ratio in QSO Absorption Systems: D/H toward 
Q1243+3047, ApJS, 149, 1-28 (2003) 
\item{} Nicastro et al., A Warm-Hot Intergalactic Medium Location for the 
Missing Cosmic Baryons, Nature, accepted for publication (2004): N04a 
\item{} Nicastro, F. et al., Chandra Detection of the First X-ray Forest 
along the Line of Sight To Mrk~421, ApJ, submitted (2004): N04b 
\item{} Nicastro, F. et al., The Far Ultraviolet Signature of the 'Missing' 
Baryons in the Local Group of Galaxies, Nature, 421, 719-721 (2003) 
\item{} Nicastro, F. et al., Chandra Discovery of a Tree in the X-ray Forest 
toward PKS~2155-304: The Local Filament?, ApJ, 573, 157-167 (2002) 
\item{} Penton, S.V., Stocke, J.T. \& Shull, J.M., The Local Ly$\alpha$ 
Forest. IV. Space Telescope Imaging Spectrograph G140M Spectra and Results on 
the Distribution and Baryon Content of H\,I Absorbers, ApJS, 152, 29-62 (2004) 
\item{} Rauch, M., The Lyman Alpha Forest in the Spectra of QSOs, ARA\&A, 36, 
267-316 (1998) 
\item{} Savage, B.D. et al., Far Ultraviolet Spectroscopic Explorer and 
Space Telescope Imaging Spectrograph Observations of Intervening O\,VI 
Absorption Line Systems in the Spectrum of PG~0953+415, ApJ, 564, 631-649 
(2002) 
\item{} Sembach, K.R. et al., Highly Ionized High-Velocity Gas in the 
Vicinity of the Galaxy, ApJS, 146, 165-208 (2003) 
\item{} Spergel, D.N. et al., First Year Wilkinson Microwave Anisotropy 
Probe (WMAP) Observations: Determination of Cosmological Parameters, ApJS, 
148, 175-194 (2003) 
\item{} Tripp, T.M., Savage, B.D. \& Jenkins, E.B., Intervening O\,VI Quasar 
Absorption Systems at Low Redshift: a Significant Baryon Reservoir, ApJ, 
534, L1-L5 (2000) 
\item{} Weinberg, D.H. et al., Lower Bound on the Cosmic Baryon Density, 
ApJ, 490, 564-570 (1997) 
\item{} Yoshikawa, K., et al.,Detectability of the Warm/Hot Intergalactic 
Medium through Emission Lines of O VII and O VIII, PASJ, 55, 879-890 (2003) 
\endthebibliography{}

\end{document}